\documentclass[11pt]{article}
\usepackage{amsmath,amssymb,amsthm,amsxtra,overpic,bbm,bm,epsfig}
\usepackage{color,subfigure}

\def\thefootnote{\fnsymbol{footnote}}

\newcommand{\bq}{\begin{eqnarray}}
\newcommand{\nq}{\end{eqnarray}}
\newcommand{\nub}{\overline{\nu}}
\newcommand{\epsilont}{\tilde{\epsilon}}
\newcommand{\depsilon}{\delta\epsilon}
\newcommand{\dvarepsilon}{\delta\varepsilon}
\newcommand{\Ut}{\tilde{U}}
\newcommand{\dU}{\delta{U}}
\newcommand{\Nt}{\tilde{N}}

\newcommand{\cost}{\tilde{c}}
\newcommand{\sint}{\tilde{s}}
\newcommand{\thetat}{\tilde{\theta}}
\newcommand{\deltat}{\tilde{\delta}}
\newcommand{\alphat}{\tilde{\alpha}}
\newcommand{\phit}{\tilde{\phi}}
\newcommand{\rhot}{\tilde{\rho}}
\newcommand{\sigmat}{\tilde{\sigma}}
\newcommand{\Pt}{\tilde{P}}
\newcommand{\eb}{\overline{e}}
\newcommand{\Deltat}{\tilde{\Delta}}
\newcommand{\Phit}{\tilde{\Phi}}
\newcommand{\At}{\tilde{A}}
\newcommand{\St}{\tilde{S}}

\newcommand{\alphab}{\overline{\alpha}}
\newcommand{\betab}{\overline{\beta}}
\newcommand{\gammab}{\overline{\gamma}}

\begin{document}

\begin{center}
{\Large\bf Shifts of neutrino oscillation parameters in reactor
antineutrino experiments with non-standard interactions}
\end{center}

\vspace{0.2cm}

\begin{center}
{\bf Yu-Feng Li} \footnote{E-mail: liyufeng@ihep.ac.cn} \quad {\bf
Ye-Ling Zhou} \footnote{E-mail: zhouyeling@ihep.ac.cn}
\\
{Institute of High Energy Physics, Chinese Academy of
Sciences, P.O. Box 918, Beijing 100049, China}
\end{center}

\vspace{1.5cm}

\begin{abstract}

We discuss reactor antineutrino oscillations with non-standard
interactions (NSIs) at the neutrino production and detection
processes. The neutrino oscillation probability is calculated with a
parametrization of the NSI parameters by splitting them into the
averages and differences of the production and detection processes
respectively. The average parts induce constant shifts of the
neutrino mixing angles from their true values, and the difference
parts can generate the energy (and baseline) dependent corrections
to the initial mass-squared differences. We stress that only the
shifts of mass-squared differences are measurable in reactor
antineutrino experiments. Taking Jiangmen Underground Neutrino
Observatory (JUNO) as an example, we analyze how NSIs influence the
standard neutrino measurements and to what extent we can constrain
the NSI parameters.

\end{abstract}

\begin{flushleft}
\hspace{0.8cm} PACS number(s): 14.60.Pq, 13.10.+q, 25.30.Pt \\
\hspace{0.8cm} Keywords: non-standard interaction, reactor
antineutrino, JUNO
\end{flushleft}

\def\thefootnote{\arabic{footnote}}
\setcounter{footnote}{0}

\newpage

\section{Introduction}
 After the observation of non-zero $\theta_{13}$ from
recent reactor \cite{DYB,DYBcpc,DYBshape,DC,RENO} and accelerator
\cite{T2K,MINOS} neutrino experiments, we have established a
standard picture of three active neutrino oscillations with three
mixing angles and two independent mass-squared differences
\cite{PDG}. Therefore the remaining neutrino mass ordering and
CP-violating phase, which manifest themselves as the generic
properties of three neutrino oscillations, constitute the main focus
of future neutrino oscillation experiments. On the other hand, the
probe of new physics beyond the Standard Model (SM) is another motivation
for future precision oscillation measurements. 

Experiments using reactor antineutrinos have played important roles
in the history of neutrino physics, which can be traced back to the
discovery of neutrinos \cite{Cowan1956}, to establishment
\cite{Kam2002} of the Large Mixing Angle (LMA)
Mikheyev-Smirnov-Wolfenstein (MSW) solution of the long-standing
solar neutrino problem, and more recently to the discovery of
non-zero $\theta_{13}$ \cite{DYB,DYBcpc,DC,RENO}. Moreover, future
reactor experiments would keep their competitive roles in the
determination of the neutrino mass hierarchy, precision measurement
of oscillation parameters, and search for additional neutrino types
and interactions. The survival probability for the reactor
antineutrino $\nub_e\to\nub_e$ oscillation in the three neutrino
framework can be written as \bq P_{\eb\eb}
=1&-&c^4_{13}\sin^22\theta_{12}\sin^2\Delta_{21}
-c^2_{12}\sin^22\theta_{13}\sin^2\Delta_{31}
-s^2_{12}\sin^22\theta_{13}\sin^2\Delta_{32}\,, 
\label{eq:PeeStandard} \nq with $c_{ij}=\cos\theta_{ij}$,
$s_{ij}=\sin\theta_{ij}$, and $\Delta_{ji}=\Delta m^2_{ji}L/(4E)$
where $L$ is the baseline distance between the source and detector,
$E$ is the antineutrino energy, and $\Delta m^2_{ji} =
m^2_{j}-m^2_{i}$ is the mass-squared difference between the $i$th
and $j$th mass eigenstates. Because there is a large hierarchy between
different mass-squared differences, \bq 30\Delta m^2_{21}\sim
|\Delta m^2_{31}| \sim |\Delta m^2_{32}|\,, \label{eq:hierarchy}\nq
different reactor antineutrino experiments may measure different
oscillation terms of $\Delta_{12}$ or ($\Delta_{31}$,
$\Delta_{32}$), which can be categorized into three different
groups:
\begin{itemize}

\item Long baseline reactor antineutrino experiments, such as KamLAND
\cite{Kam2002,Kam2011}. The aim of these experiments is to observe
the slow oscillation with $\Delta_{21}$ and measure the
corresponding oscillation parameters $\Delta m^2_{21}$ and
$\theta_{12}$.

\item Short baseline reactor antineutrino experiments, such
as Daya Bay \cite{DYB,DYBcpc,DYBshape}, Double CHOOZ \cite{DC}, RENO
\cite{RENO}. They are designed to observe the fast oscillation with
$\Delta_{31}$ and $\Delta_{32}$ (or equivalently, $\Delta_{ee}$
\cite{DYBshape}) and measure the corresponding oscillation
parameters $\Delta m^2_{ee}$, $\theta_{13}$.

\item Medium baseline reactor antineutrino experiments. They stand for
the next generation experiments of reactor antineutrinos, with
typical representatives of Jiangmen Underground Neutrino Observatory
(JUNO) \cite{Li:2013zyd} and RENO-50 \cite{RENO50}. They can
determine the neutrino mass ordering ($m_1<m_2<m_3$ or
$m_3<m_1<m_2$). In addition, they are expected to provide the
precise measurement for both the fast and slow oscillations and
become a bridge between short baseline and long baseline reactor
antineutrino experiments.

\end{itemize}

High-dimensional operators originating from new physics can
contribute to the neutrino oscillation in the form of non-standard
interactions (NSIs) \cite{Antusch:2008tz,NSIorigin}. They induce
effective four-fermion interactions after integrating out some heavy
particles beyond the SM, where the heavy particles can be scalars,
pseudo-scalars, vectors, axial-vectors, or tensors
\cite{Severijns:2006dr}. For reactor antineutrino experiments NSIs
may appear in the antineutrino production and detection processes,
and can modify the neutrino oscillation probability. Therefore, the
neutrino mixing angles and mass-squared differences can be shifted
and the mass ordering (MO) measurement will be affected. There are
some previous discussions on NSIs in reactor antineutrino
experiments \cite{Ohlsson:2008gx,Ohlsson:2013nna, Khan:2013hva} and
other types of oscillation experiments \cite{generalNSI}. In this
work, we study the NSI effect in reactor antineutrino oscillations
in both specific models and also the most general case. Taking JUNO
as an example, we apply our general framework to the medium baseline
reactor antineutrino experiment. We discuss how NSIs influence the
standard 3-generation neutrino oscillation measurements and to what
extent we can constrain the NSI parameters.

The remaining part of this work is organized as follows. Section 2
is to derive the analytical formalism. We develop a general
framework on the NSI effect in reaction antineutrino oscillations,
and calculate the neutrino survival probability in the presence of
NSIs. In section 3, we give the numerical analysis for the JUNO
experiment. We analyze the NSI impacts on the precision measurement
of mass-squared differences and the determination of the neutrino
mass ordering, and present the JUNO sensitivity of the relevant NSI
parameters. Finally, we conclude in section 4.

\section{NSI-induced neutrino oscillations}

\subsection{Basic formalism}
\label{subsec:formula}
NSIs may occur in the neutrino production,
detection and propagation processes in neutrino oscillation
experiments. The neutrino and antineutrino states produced in the
source and observed in the
detector are superpositions of flavor states,
\bq &&|\nu^{\rm s}_\alpha\rangle = \frac{1}{N^{\rm s}_\alpha}
\Big(|\nu_\alpha\rangle+\sum_{\beta}
\epsilon^{\rm s}_{\alpha\beta}|\nu_\beta\rangle \Big),\qquad
|\nub^{\rm s}_\alpha\rangle = \frac{1}{N^{\rm s}_\alpha} \Big(|\nub_\alpha\rangle+\sum_{\beta}
\epsilon^{\rm s*}_{\alpha\beta}|\nub_\beta\rangle \Big), \nonumber\\
&&\langle\nu^{\rm d}_\beta| = \frac{1}{N^{\rm d}_\beta}
\Big(\langle\nu_\beta|+\sum_{\alpha}
\epsilon^{\rm d}_{\alpha\beta}\langle\nu_\alpha| \Big), \qquad
\langle\nub^{\rm d}_\beta| = \frac{1}{N^{\rm d}_\beta}
\Big(\langle\nub_\beta|+\sum_{\alpha}
\epsilon^{\rm d*}_{\alpha\beta}\langle\nub_\alpha| \Big),\nq
in which the superscripts `s' and `d' denote the source and
detector, respectively, and \bq &N^{\rm s}_\alpha=\sqrt{\sum_\beta
|\delta_{\alpha\beta}+\epsilon^{\rm s}_{\alpha\beta}|^2}, \qquad
N^{\rm d}_\beta=\sqrt{\sum_\alpha
|\delta_{\alpha\beta}+\epsilon^{\rm d}_{\alpha\beta}|^2}
\label{eq:normalize_factor_sd} \nq
are normalization factors.

In general, NSIs in different physical processes may have
distinct contributions. For a certain type of neutrino experiments,
the same set of effective NSI parameters can be introduced to
describe the NSI effect. But when one turns to anther type of
neutrino experiments, neutrinos can have totally different origins
and one should use another set of NSI effective parameters to
parametrize the NSI effect. The parameters
$\epsilon^{\rm s,d}_{\alpha\beta}$ used here are strongly
experiment-dependent, and in principle also energy-dependent.
However, they are usually considered as the averaged effects and
treated as constant values.

In order to measure the average and difference between neutrino
production and detection processes, we introduce two sets of NSI
parameters as
\bq
&&\epsilont_{\alpha\beta}=(\epsilon^{\rm s}_{\alpha\beta}+\epsilon^{\rm d*}_{\beta\alpha})/2\,,\qquad
\depsilon_{\alpha\beta}=(\epsilon^{\rm s}_{\alpha\beta}-\epsilon^{\rm d*}_{\beta\alpha})/2\,,
\label{eq:definition} \nq
to rewrite the NSI effect. One should note that the NSI parameters
$\delta\epsilon_{\alpha\beta}$ are negligibly small compared with
$\epsilont_{\alpha\beta}$ when neutrinos are purely left-handed
particles \cite{Severijns:2006dr,generalNSI}. However, if neutrinos
are right-handed particles and the four-fermion interactions are
mediated by heavy scalars beyond SM, $\depsilon_{e\alpha}$ can reach
the percent level \cite{Severijns:2006dr}.

The effective Hamiltonian that describes the vacuum neutrino
oscillation is given by
\bq H=\frac{1}{2E}\left[ U^* \text{diag}(m_1^2, m_2^2, m_3^2) U^T
\right]\,, \label{eq:vachamiltonian} \nq
where $U$ is the Pontecorvo-Maki-Nakagawa-Sakata (PMNS) matrix \cite{PMNS} and can be expressed in the form \cite{Mei:2003gn}
\bq U=P_l^\dag
\begin{pmatrix}
c_{12}c_{13} & s_{12}c_{13} & s_{13} \\
-s_{12}c_{23}e^{-i\delta}-c_{12}s_{23}s_{13} &
\hspace{8pt}c_{12}c_{23}e^{-i\delta}-s_{12}s_{23}s_{13} &
s_{23}c_{13} \\
\hspace{8pt}s_{12}s_{23}e^{-i\delta}-c_{12}c_{23}s_{13} &
-c_{12}s_{23}e^{-i\delta}-s_{12}c_{23}s_{13} & c_{23}c_{13}
\end{pmatrix}
P_\nu, \label{eq:PMNS} \nq
where $s_{ij}=\sin\theta_{ij}$, $P_l={\rm
diag}\{e^{i\phi_1},e^{i\phi_2},e^{i\phi_3}\}$, and $P_\nu={\rm
diag}\{1,e^{i\rho},e^{i\sigma}\}$. $P_l$ is unphysical since charged
leptons are Dirac particles. On the other hand, $P_\nu$ can not be
neglected for Majorana neutrinos but does not contribute to neutrino
oscillations. We write the PMNS matrix in the form as in Eq.~\eqref{eq:PMNS} to keep the first row and the third column of $P_l U P^\dag_\nu$ real. The mixing angles $\theta_{12}$, $\theta_{13}$, $\theta_{23}$ and CP phase $\delta$ take the same values as those in the PDG parametrization \cite{PDG}, respectively.

In the presence of NSI effects at the source and detector, the
amplitude of the $\nub_\alpha\to\nub_\alpha$ transition is
\bq \mathcal{\At}_{\alphab\alphab}&=&\langle \nub^{\rm d}_\alpha| e^{-iHL}
|\nub^{\rm s}_\alpha \rangle= \sum_{\beta\gamma}\langle
\nub^{\rm d}_\alpha|\nub_\gamma \rangle \langle \nub_\gamma| e^{-iHL}
|\nub_\beta \rangle \langle \nub_\beta|\nub^{\rm s}_\alpha \rangle
=(\delta_{\gamma\alpha}+\epsilon^{\rm d*}_{\gamma\alpha})
\mathcal{A}_{\betab\gammab}
(\delta_{\alpha\beta}+\epsilon^{\rm s*}_{\alpha\beta}) \,,\nq
where $L$ is the baseline, and
\bq \mathcal{A}_{\betab\gammab} =\langle \nub_\gamma| e^{-iHL}
|\nub_\beta \rangle = \sum_i U^*_{\gamma i}U_{\beta i}
\exp\left(-i\frac{m^2_iL}{2E}\right) \nq
is the amplitude of $\nub_\beta\to\nub_\gamma$ without NSIs. It is
useful to define
\bq \Ut_{\alpha i}=\frac{1}{\Nt_\alpha}
\sum_\beta(\delta_{\alpha\beta}+\epsilont_{\alpha\beta}^*)U_{\beta
i}, \qquad \dU_{\alpha i}=\frac{1}{\Nt_\alpha}
\sum_\beta\depsilon_{\alpha\beta}^*U_{\beta i},
\label{eq:effectivePMNS} \nq
where
\bq &\Nt_\alpha =\sqrt{\sum_\beta
|\delta_{\alpha\beta}+\epsilont_{\alpha\beta}|^2}=\sqrt{N^{\rm s}_\alpha
N^{\rm d}_\alpha} +\mathcal{O}(\depsilon^2)\,, \label{eq:normalize_factor}
\nq
and $\sum_{i}|\Ut_{\alpha i}|^2=1$ is required. Thus we can obtain
$\mathcal{\At}_{\alphab\alphab}$ as
\bq \mathcal{\At}_{\alphab\alphab}&=& \sum_i(\Ut-\dU)^*_{\alpha
i}(\Ut+\dU)_{\alpha i} \exp\left(-i\frac{m^2_iL}{2E}\right)
+\mathcal{O}(\depsilon^2)\,, \nq
and the survival probability for $\nub_\alpha\to\nub_\alpha$ is
expressed as
\bq \Pt_{\alphab\alphab} =|\mathcal{\At}_{\alphab\alphab}|^2=
1-4\sum_{i<j}|\Ut_{\alpha i}|^2|\Ut_{\alpha j}|^2\ \left[
\sin^2\Delta_{ji} +{\rm Im}\left(\frac{\dU_{\alpha i}}{\Ut_{\alpha
i}}-\frac{\dU_{\alpha j}}{\Ut_{\alpha j}}\right) \sin2\Delta_{ji}
\right] +\mathcal{O}(\depsilon^2)\,.\nq
Because of the smallness of $\dU_{\alpha i}/\Ut_{\alpha i}$, we can
rewrite the above equation as
\bq \Pt_{\alphab\alphab} = 1-4\sum_{i<j}|\Ut_{\alpha
i}|^2|\Ut_{\alpha j}|^2\ \sin^2\Deltat_{ji}^\alpha
+\mathcal{O}(\depsilon^2) \label{eq:palphavaccumlike} \nq
with
\bq \Deltat_{ji}^\alpha=\Delta_{ji}+{\rm Im}\left(\frac{\dU_{\alpha
i}}{\Ut_{\alpha i}}-\frac{\dU_{\alpha j}}{\Ut_{\alpha j}}\right) .
\nq
Note that
$\Deltat_{31}^\alpha=\Deltat_{32}^\alpha+\Deltat_{21}^\alpha$ holds,
just as the relation in the standard three neutrino case with
$\Delta_{31}=\Delta_{32}+\Delta_{21}$. For the effective
mass-squared differences we have
\bq \Delta \tilde{m}^{2\alpha}_{ji}({E}/{L})=\Delta m^2_{ji}+{\rm
Im}\left(\frac{\dU_{\alpha i}}{\Ut_{\alpha i}}-\frac{\dU_{\alpha
j}}{\Ut_{\alpha j}}\right)\frac{4E}{L}\,, \nq
which is an energy/baseline- and flavor-dependent effective quantity.

With Eq.~(\ref{eq:palphavaccumlike}) we have obtained a
standard-like expression for the antineutrino survival probability in
vacuum in the presence of NSIs. The corresponding NSI effects are
encoded in the effective mass and mixing parameters. The average
parts induce constant shifts for the neutrino mixing elements, and
the difference parts generate energy and baseline dependent
corrections to the mass-squared differences. Without prior
information on the true mass and mixing parameters, we cannot
distinguish between the true and effective parameters. Only the
energy- and baseline-dependent feature of the effective parameters
can tell us the NSI effects, and correspondingly the difference
parts of NSIs will be constrained with precise spectral measurements
of reactor antineutrino oscillations.

\subsection{Parametrization of the effective PMNS matrix $\Ut$}

We have defined the matrix $\Ut$ in Eq.~\eqref{eq:effectivePMNS},
which is considered as an effective PMNS matrix compared with the
original one $U$ in the standard three neutrino framework. However,
the unitary conditions for $\Ut$  do not hold any more. Instead, we
have
\bq
&&\sum_{i}|\Ut_{\alpha i}|^2=1\,, \nonumber\\
&&\sum_{i}\Ut_{\alpha i}\Ut^*_{\beta i}=\epsilont_{\alpha\beta}^*+
\epsilont_{\beta\alpha}+\mathcal{O}(\epsilont^2) \quad\text{for}~\alpha\neq\beta\,,\nonumber\\
&&\sum_{\alpha}\Ut_{\alpha i}\Ut^*_{\alpha
j}=\delta_{ij}+\sum_{\alpha,\beta}U_{\alpha i}U_{\beta j}^*
(\epsilont^*_{\alpha\beta}+\epsilont_{\beta\alpha})+\mathcal{O}(\epsilont^2)
\quad\text{for}~\alpha\neq\beta.
\label{eq:norm}
\nq
We plan to perform a similar parametrization  for $\Ut$ as in
Eq.~(\ref{eq:PMNS}). However, $\Ut$ is not a unitary matrix and to
perform such a parametrization is not a trivial task. Taking account
of the relation in Eq.~\eqref{eq:norm} and the detailed
configuration of neutrino oscillation experiments, we parametrize
$\Ut$ as follows, with $\sint_{ij}=\sin\thetat_{ij}$ and
$\cost_{ij}=\cos\thetat_{ij}$ for simplicity,
\begin{itemize}
\item[1)] We first extract two diagonal phase matrices $\Pt_l={\rm diag}\{e^{i\phit_1},e^{i\phit_2},e^{i\phit_3}\}$
and $\Pt_\nu={\rm diag}\{1,e^{i\rhot},e^{i\sigmat}\}$ to make the
first row and third column of $\Pt_l\Ut\Pt_\nu^\dag$ real and
positive. $\Pt_l$ includes unphysical phases to be redefined by
rephasing the charged lepton fields,  but $\rhot$ and $\sigmat$ are
the effective Majorana CP phases modified by NSIs, which do not
contribute to the neutrino oscillation but will essentially
influence the process of the neutrinoless double beta decay.

\item[2)] We define $\thetat_{13}$ and $\thetat_{12}$ through $\sint_{13}=|\Ut_{e3}|$ and $\sint_{12}=|\Ut_{e2}|/\cost_{13}$.
Using the normalization relation
$|\Ut_{e1}|^2+|\Ut_{e2}|^2+|\Ut_{e3}|^2=1$, we have
$|\Ut_{e1}|=\cost_{12}\cost_{13}$. These two mixing angles are
directly related to the reactor antineutrino oscillation
experiments.

\item[3)] Using $\thetat_{13}$ defined in 2), we define $\sint_{23}=|\Ut_{\mu3}|/\cost_{13}$, with the equation
$|\cost_{12}\cost_{23}e^{-i\deltat}-\sint_{12}\sint_{23}\sint_{13}|=|\Ut_{\mu2}|$,
we obtain a definition of the CP-violating phase $\deltat$.
$|\Ut_{\mu1}|=|-\sint_{12}\cost_{23}e^{-i\deltat}-\cost_{12}\sint_{23}\sint_{13}|$
is defined from the normalization relation
$|\Ut_{\mu1}|^2+|\Ut_{\mu2}|^2+|\Ut_{\mu3}|^2=1$.

\item[4)] Since $|\Ut_{e3}|^2+|\Ut_{\mu3}|^2+|\Ut_{\tau3}|^2\neq1$, a fourth mixing angle $\thetat'_{23}$ is needed
via the definition of $\cost'_{23}=|\Ut_{\tau3}|/\cost_{13}$. In
addition, we define a second CP-violating phase $\deltat'$ from the
equation
$|-\cost_{12}\sint'_{23}e^{-i\deltat'}-\sint_{12}\cost'_{23}\sint_{13}|=|\Ut_{\tau2}|$.
Meanwhile,
$|\Ut_{\tau1}|=|\sint_{12}\sint'_{23}e^{-i\deltat'}-\cost_{12}\cost'_{23}\sint_{13}|$
is obtained from the normalization relation
$|\Ut_{\tau1}|^2+|\Ut_{\tau2}|^2+|\Ut_{\tau3}|^2=1$.
\item[5)] Finally, there are 4 additional CP-violating phases
$\alphat_{\mu1}$, $\alphat_{\mu2}$, $\alphat_{\tau1}$ and
$\alphat_{\tau2}$ which are defined as follows:
\bq
\alphat_{\mu1}=\arg\left(\frac{\Ut_{\mu1}\Ut_{e3}}{\Ut_{\mu3}\Ut_{e1}}\right),
\alphat_{\mu2}=\arg\left(\frac{\Ut_{\mu2}\Ut_{e3}}{\Ut_{\mu3}\Ut_{e2}}\right),
\alphat_{\tau1}=\arg\left(\frac{\Ut_{\tau1}\Ut_{e3}}{\Ut_{\tau3}\Ut_{e1}}\right),
\alphat_{\tau2}=\arg\left(\frac{\Ut_{\tau2}\Ut_{e3}}{\Ut_{\tau3}\Ut_{e2}}\right).
\nq
\end{itemize}
To summarize, we have obtained a parametrization similar to Eq.~\eqref{eq:PMNS}:
\bq \Ut=\Pt_l^\dag
\begin{pmatrix}
\cost_{12}\cost_{13} & \sint_{12}\cost_{13} & \sint_{13} \\
(-\sint_{12}\cost_{23}e^{-i\deltat}-\cost_{12}\sint_{23}\sint_{13})e^{i\alphat_{\mu1}}
&
\hspace{10pt}(\cost_{12}\cost_{23}e^{-i\deltat}-\sint_{12}\sint_{23}\sint_{13})e^{i\alphat_{\mu2}}
&
\sint_{23}\cost_{13} \\
\hspace{6pt}(\sint_{12}\sint'_{23}e^{-i\deltat'}-\cost_{12}\cost'_{23}\sint_{13})e^{i\alphat_{\tau1}}
&
(-\cost_{12}\sint'_{23}e^{-i\deltat'}-\sint_{12}\cost'_{23}\sint_{13})e^{i\alphat_{\tau2}}
& \cost'_{23}\cost_{13}
\end{pmatrix}
\Pt_\nu. \label{eq:reparametrization} \nq
Fifteen parameters are included in $\Ut$ and ten of them
($\thetat_{12},~\thetat_{13},~\thetat_{23},~\thetat'_{23},~\deltat,~\deltat',~\alphat_{\mu1},~\alphat_{\mu2},
~\alphat_{\tau1}, ~\alphat_{\tau2}$) are related to neutrino
oscillations. In the case of $\epsilont\to 0$, we can restore the
original PMNS matrix of three active neutrinos as
 \bq
\thetat_{13}\to\theta_{13},\qquad \thetat_{12}\to\theta_{12},\qquad
\thetat_{23},\thetat'_{23}\to\theta_{23},\qquad
\deltat,\deltat'\to\delta,\qquad
\alphat_{\mu1},\alphat_{\mu2},\alphat_{\tau1},\alphat_{\tau2}\to 0.
\nq
One should notice that although the effective PMNS matrix takes a
simple form as in Eq.~\eqref{eq:reparametrization}, the
corresponding parameters are in general dependent on the type of
experiments. We should use different $\Ut$ to characterize different
realizations of the effective PMNS matrix for reactor and
accelerator neutrino experiments. This is different from the
non-unitary effect of the PMNS matrix, where $\epsilont$ is used to
parametrize the universal mixing between the active and sterile
neutrinos and $\depsilon=0$ by definition. In this case, $\Ut$ is an
effective PMNS matrix for all neutrino oscillation experiments.

\subsection{Reactor antineutrino oscillation probabilities}

In reactor antineutrino oscillations, only the electron antineutrino
survival probability is relevant because of the high threshold of
the $\mu$/$\tau$ production. With the parametrization of $\Ut$ in
Eq.~\eqref{eq:reparametrization}, we can rewrite $P_{\eb\eb}$ with
these effective mixing parameters as
\bq
\Pt_{\eb\eb}=1&-&\cost^4_{13}\sin^22\thetat_{12}[\sin^2\Delta_{21}+(\dvarepsilon_1-\dvarepsilon_2)\sin2\Delta_{21}] \nonumber\\
&-&\cost^2_{12}\sin^22\thetat_{13}[\sin^2\Delta_{31}+(\dvarepsilon_1-\dvarepsilon_3)\sin2\Delta_{31}]\nonumber\\
&-&\sint^2_{12}\sin^22\thetat_{13}[\sin^2\Delta_{32}+(\dvarepsilon_2-\dvarepsilon_3)\sin2\Delta_{32}]
+\mathcal{O}(\depsilon^2) \nonumber\\
=1&-&\cost^4_{13}\sin^22\thetat_{12}\sin^2\Deltat_{21}
-\cost^2_{12}\sin^22\thetat_{13}\sin^2\Deltat_{31}
-\sint^2_{12}\sin^22\thetat_{13}\sin^2\Deltat_{32}
+\mathcal{O}(\dvarepsilon^2) \label{eq:reactorprobability} \nq
with
\bq
&&\Deltat_{ji}=\Delta_{ji}+\dvarepsilon_i-\dvarepsilon_j, \nonumber\\
&&\dvarepsilon_1=\frac{{\rm
Im}(\dU_{e1})}{\cost_{12}\cost_{13}},\quad \dvarepsilon_2=\frac{{\rm
Im}(\dU_{e2})}{\sint_{12}\cost_{13}},\quad \dvarepsilon_3=\frac{{\rm
Im}(\dU_{e3})}{\sint_{13}}\,,
\label{eq:epsilon_definition}
\nq
where the superscript $\alpha=e$ in $\Deltat_{ji}^e$ has been
ignored.

The average part $\epsilont$ can be treated as constant shifts to
mixing angles $\theta_{12}$ and $\theta_{13}$, and the difference
part $\depsilon$ leads to energy- and baseline-dependent shifts to
the mass-squared differences $\Delta m^2_{ji}$\, as
\bq \Delta \tilde{m}^2_{ji}(E/L)=\Delta
m^2_{ji}+(\dvarepsilon_i-\dvarepsilon_j)4E/L\,.
\label{eq:mass_shift} \nq
However, only two combinations of the three parameters
$\dvarepsilon_i$ contribute to the oscillation probability thanks to
the relation
$(\dvarepsilon_2-\dvarepsilon_3)=(\dvarepsilon_1-\dvarepsilon_3)-(\dvarepsilon_1-\dvarepsilon_2)$.
It is notable that one cannot distinguish the effect of mixing angle
shifts from the scenario of three neutrino mixing using reactor
antineutrino oscillations. This degeneracy can only be resolved by
including different types of neutrino oscillation experiments, where
the NSI parameters and their roles in neutrino oscillations are
totally distinct. On the other hand, the shifts of mass-squared
differences are clearly observable due to the baseline- and
energy-dependent corrections in the reactor antineutrino spectrum.

Different kinds of reactor antineutrino oscillation experiments have their own advantages
in measuring the NSI parameters. The long baseline reactor
antineutrino experiment (e.g., KamLAND) can measure the slow
oscillation term $\Delta_{21}$ and thus are sensitive to the
measurement of $\dvarepsilon_1-\dvarepsilon_2$. Since the fast
oscillation terms $\Delta_{31}$ and $\Delta_{32}$ are averaged out, the oscillation
probability $\Pt_{\eb\eb}$ is reduced to
\bq \Pt_{\eb\eb}=\sint^4_{13}+\cost^4_{13}\big\{
1-\sin^22\thetat_{12} \big[
\sin^2\Delta_{21}+(\dvarepsilon_1-\dvarepsilon_2)\sin2\Delta_{21}
\big] \big\}\,. \label{eq:PeeKL} \nq
The short baseline reactor antineutrino experiments (e.g., Daya Bay)
are designed to measure the fast oscillation terms $\Delta_{31}$ and
$\Delta_{32}$ (or equivalently, $\Delta_{ee}$), and are 
effective to constrain $\dvarepsilon_1-\dvarepsilon_3$. Since the
slow oscillation term $\Delta_{21}$ is negligible, the oscillation
probability $\Pt_{\eb\eb}$ can be simplified as
\bq
\Pt_{\eb\eb}=1-\sin^22\thetat_{13}\sin^2\Delta_{ee}-\eta
\Big[\cost^2_{12}(\dvarepsilon_1-\dvarepsilon_3)+\sint^2_{12}(\dvarepsilon_2-
\dvarepsilon_3)\Big]\sin^22\thetat_{13}\sin2\Delta_{ee}.
\label{eq:PeeDYB} \nq
For the reactor antineutrino oscillation at the medium baseline (e.g.,
$52.5\,$km), we can generalize the formalism given in
Ref.~\cite{Minakata:2007tn} to see how the mass ordering measurement
can be influenced by the NSI effect. The oscillation probability in
Eq.~\eqref{eq:reactorprobability} can be rewritten as
\bq
\Pt_{\eb\eb}=1&-&\frac{1}{2}\sin^22\thetat_{13}-\cost^4_{13}\sin^22\thetat_{12}
[\sin^2\Delta_{21}+(\dvarepsilon_1-\dvarepsilon_2)\sin2\Delta_{21}]\nonumber\\
&+&\frac{1}{2}\sin^22\thetat_{13}(C\cos2\Delta_{ee}-\eta
S\sin2\Delta_{ee}), \label{eq:Pee} \nq
where
\begin{eqnarray}
\Delta_{ee}= \frac{\Delta m^2_{ee}L}{4E}\,,\qquad
\Delta m^2_{ee}= \cost^2_{12}|\Delta m^2_{31}| + \sint^2_{12}|\Delta
m^2_{32}|\,, \label{dmee}
\end{eqnarray}
and
\bq
C&=&\cost^2_{12}\cos(2\sint^2_{12}\Delta_{21})+\sint^2_{12}\cos(2\cost^2_{12}\Delta_{21})\nonumber\\
&-&2(\dvarepsilon_1-\dvarepsilon_3)\cost^2_{12}\sin(2\sint^2_{12}\Delta_{21})
+2(\dvarepsilon_2-\dvarepsilon_3)\sint^2_{12}\sin(2\cost^2_{12}\Delta_{21}), \nonumber\\
S&=&\cost^2_{12}\sin(2\sint^2_{12}\Delta_{21})-\sint^2_{12}\sin(2\cost^2_{12}\Delta_{21})\nonumber\\
&+&2(\dvarepsilon_1-\dvarepsilon_3)\cost^2_{12}\cos(2\sint^2_{12}\Delta_{21})
+2(\dvarepsilon_2-\dvarepsilon_3)\sint^2_{12}\cos(2\cost^2_{12}\Delta_{21}),
\nq
with $\eta=\pm1$ for the normal mass ordering (NMO) and inverted
mass ordering (IMO), respectively. As a function of
$\dvarepsilon_1-\dvarepsilon_3$ and $\dvarepsilon_2-\dvarepsilon_3$,
$S$ varies with the NSI parameters, which can further alter the
difference of oscillation probabilities between NMO and IMO. Since
NSIs are constrained to the percent level \cite{Biggio:2009nt}, we
expect that NSIs will not significantly affect the mass ordering
measurement.

Finally, one should keep in mind that there is an additional correction from
terrestrial matter effects during the neutrino propagation inside
the Earth. In addition to the Hamiltonian in
Eq.~(\ref{eq:vachamiltonian}), there is a matter potential term
written as
\bq H_{\rm mat}=\frac{1}{2E}\text{diag}(-A_{\rm CC},0,0)\,, \nq
where $-A_{\rm CC}$ characterizes the contribution of charged-current
interactions between antineutrinos and electrons in matter and
$A_{\rm CC}=2\sqrt{2}G_F N_e E$,
with $N_e$ being the electron number density. For reactor
antineutrino experiments, $A_{\rm CC}$ is sufficiently small
compared with the kinetic term $\Delta m^2_{ji}$ where the matter
corrections to oscillation parameters of the solar sector are given
by
\bq
\sin^22\theta_{12}^{\text{M}}&\simeq&\sin^22\theta_{12}(1-
2\frac{A_{\text{CC}}}{\Delta {m}^2_{21}}\cos2\theta_{12})\,, \nonumber\\
\Delta {m}^{2\text{M}}_{21} &\simeq&\Delta
{m}^2_{21}(1+\frac{A_{\text{CC}}}{\Delta
{m}^2_{21}}\cos2\theta_{21}) \,,\nq
and the magnitude of these corrections is estimated as
 \bq \frac{A_{\text{CC}}}{\Delta
{m}^2_{21}}\cos2\theta_{12}\simeq 0.5\% \times
\frac{E}{4\text{MeV}}\times \frac{\rho}{3\text{g}/\text{cm}^3}\,,
\nq
with $\rho$ being the matter density along the antineutrino
trajectory in the Earth. In comparison, the correction to parameters
of the atmospheric sector is only the order of $10^{-5}$. In our
following studies, matter effects during the neutrino propagation
will be neglected in the analytical calculation but effectively
included in our numerical analysis.

The $N_e E$ suppression leads to negligible NSI effects in the
propagation process. The scenario is very different from other types
of neutrino oscillation experiments, where NSIs during propagation
lead to much larger corrections to oscillation probabilities than
those at the source and detector. In reactor antineutrino
oscillations NSIs during propagation can be safely neglected.

\section{Numerical simulation}
In this section, following the JUNO nominal setup in
Ref.~\cite{Li:2013zyd}, we will numerically show the NSI effect in
the medium baseline reactor antineutrino experiment. We will
illustrate how the NSI effect shifts the mass-squared differences,
how it influences the mass ordering measurement and to what extent
we can constrain the NSI parameters.

\subsection{Nominal setups}
\label{subsec:setups}
We employ the true power and baseline
distribution in Table 1 of Ref.~\cite{Li:2013zyd}. The weighted
average of the baseline is around 52.5 km with baseline differences
less than 500 m. We use the nominal running time of 1800 effective
days for six years, and detector energy resolution
$3\%/\sqrt{E(\text{MeV})}$ as a benchmark. A NMO is assumed to be
true otherwise mentioned explicitly. The relevant oscillation
parameters are $\thetat_{12}$, $\thetat_{13}$, $\Delta m^2_{21}$ and
$\Delta m^2_{ee}$ and the NSI parameters are $\dvarepsilon_1-\dvarepsilon_2$, $\dvarepsilon_1-\dvarepsilon_3$.

We directly employ the mixing angles measured in recent reactor
antineutrino experiments as our effective mixing angles, which can
be shown as
\bq \sin^22\thetat_{13}=\sin^22\theta_{13}^\text{D}=0.084, \quad
\tan^2\thetat_{12}=\tan^2\theta_{12}^\text{K}=0.481.
\label{eq:datamixing} \nq
The measured mixing angles $\theta_{13}^\text{D}$ and
$\theta_{12}^\text{K}$ are from Daya Bay \cite{An:2013zwz} and
KamLAND \cite{Gando:2013nba}, respectively. For some recent
discussions on how the true mixing angles are modified by the
average of the NSI effect in a simplified version, see
\cite{Ohlsson:2013nna,Girardi:2014gna}. However, the measured
mass-squared differences $\Delta m^{2\,\text{D}}_{ee}$ from Daya Bay
\cite{An:2013zwz} and $\Delta m^{2\,\text{K}}_{21}$ from KamLAND
\cite{Gando:2013nba} can be viewed as the true parameters rather
than effective oscillation parameters, i.e.,
\bq \Delta m^2_{ee}=\Delta
m^{2\,\text{D}}_{ee}=2.44\times10^{-3}\,\text{eV}^2, \quad \Delta
m^2_{21}=\Delta m^{2\,\text{K}}_{21}=7.54\times10^{-5}\,\text{eV}^2.
\label{eq:datamass} \nq
The reason is that although the mass-squared differences have
energy/baseline-dependent corrections as shown in
Eq.~\eqref{eq:mass_shift}, these shifts are sufficiently small
compared with the current level of uncertainties. In details,
$\dvarepsilon_i$ may be in the percent level and the ratio $E/L$ for
Daya Bay is around $(E/L)^\text{D}\simeq2
~\text{MeV/km}\simeq4\times10^{-4}~\text{eV}^2$, and for KamLAND is
around $(E/L)^\text{K}\simeq0.02
~\text{MeV/km}\simeq4\times10^{-6}~\text{eV}^2$. Therefore the
shifts are still below the sensitivity of Daya Bay \cite{An:2013zwz}
and KamLAND \cite{Gando:2013nba}, and the measured parameters can be
approximate to the true values for $\Delta m^2_{ee}$ and $\Delta
m^2_{21}$, respectively.

In our following numerical analysis, two different treatments for
the NSI parameters will be explored.
\begin{itemize}
\item The first treatment is a class
of specific models with democratic entries for
$\depsilon_{\alpha\beta}$, and for simplicity we assume that these
models have identical magnitude in ${\rm Im}(\dU_{e i})$ but
different relative signs. The configurations of
$\{\text{Im}(\dU_{e1}),\text{Im}(\dU_{e2}),\text{Im}(\dU_{e3})\}$
for these specific models are defined as
 \bq &&{\rm S1}:~(\dU_{},+\dU_{},
+\dU_{})\,,\quad\, {\rm
S2}:~(\dU_{},+\dU_{}, -\dU_{})\,,\nonumber\\
&&{\rm S3}:~(\dU_{}, -\dU_{}, +\dU_{})\,,\quad\, {\rm
S4}:~(\dU_{}, -\dU_{}, -\dU_{})\,, \label{eq:specmod}
\nq
respectively. Therefore from Eq.~\eqref{eq:epsilon_definition}, $\dvarepsilon_2$ can be roughly larger than
$\dvarepsilon_1$ due to $\theta_{12}<45^\circ$ and both of them are
the same order of $\dU_{}$. $\dvarepsilon_3$ can be several times
larger due to the smallness of $\theta_{13}$.

\item The second one is the general treatment for the NSI parameters. As
shown in Eq.~(\ref{eq:reactorprobability}), only two combinations of
the NSI parameters $\dvarepsilon_i$ are independent, thus we can
treat $(\dvarepsilon_1-\dvarepsilon_2)$ and
$(\dvarepsilon_1-\dvarepsilon_3)$ as free parameters to cover the
full parameter space. Note that the non-unitary effect is identical
with NSIs for the case of $\depsilon_{\alpha\beta}=0$. If we can
observe the splittings of $\Deltat_{ji}^\alpha$ compared with
$\Delta_{ji}$, we may distinguish NSIs from the non-unitary effect.
\end{itemize}

\subsection{Antineutrino spectrum}
To calculate the expected reactor $\nub_e$ spectrum in the presence
of NSIs, we need first deal with the standard case without the NSI
effect. The energy spectrum of detected events $S(E_\text{vis})$, as
a function of the visible energy $E_\text{vis}$ of the inverse
$\beta$-decay $\nub_e+p\to e^++n$ (IBD), is parametrized as
\bq S(E_\text{vis}) = \int_{m_e}^\infty dE_e \left[\int_{E_T}^\infty
dE \Big(\sum_i \;{\cal N}_i \;\Phi_i(E) P_{\eb\eb}(E/L_i)\Big)
\frac{d\sigma(E,\,E_e)}{dE_e}\right]r(E_e+m_e,\,E_\text{vis})
, \label{eq:spectrum} \nq
where $\Phi_i(E)$ is the antineutrino flux with $i$ standing for
different reactor cores and $E$ the antineutrino energy, ${\cal
N}_i$ is the corresponding normalization and conversion factor,
$P_{\eb\eb}(E/L_i)$ is the oscillation probability of
$\nub_e\to\nub_e$ with different baseline $L_i$ from the $\nub_e$
source $i$ to the detector, $d\sigma(E,E_e)/dE_e$ is the IBD differential cross section
with $E_e$ being the true positron energy,
$r(E_e+m_e,\,E_\text{vis})$ is the Gaussian energy resolution
function with a standard deviation $\sigma_{E}$ defined as
\bq \frac{\sigma_{E}}{E_e+m_e}=\frac{3\%}{\sqrt{(E_e+m_e)/\rm
MeV}}\,. \label{eq:Eres} \nq
In the presence of NSIs, we can use the following replacements to
show the NSI effect during the antineutrino oscillation, production and detection
processes, respectively \bq P_{\eb\eb}\to \Pt_{\eb\eb}, \qquad
\Phi_i\to \Phit_i=(N_e^{\rm s})^2 \Phi_i, \qquad \sigma \to
\sigmat=(N_e^{\rm d})^2 \sigma. \nq Therefore, we can obtain the NSI-modified reactor $\nub_e$ spectrum as
\bq \St(E_\text{vis}) &=& (\Nt_e)^4 \int_{m_e}^\infty dE_e\left[
\int_{E_T}^\infty dE \Big(\sum_i \;{\cal N}_i \;\Phi_i(E)
\Pt_{\eb\eb}(E/L_i)\Big) \frac{d\sigma(E,\,E_e)}{dE_e}\right]
r(E_e+m_e,\,E_\text{vis})
+ \mathcal{O}(\depsilon^2)\,, \nonumber\\
\label{eq:spectrum_NSIs} \nq
where $N_e^{\rm s}$, $N_e^{\rm d}$ and $\Nt_{e}$ are defined in
Eqs.~\eqref{eq:normalize_factor_sd} and \eqref{eq:normalize_factor},
and can be absorbed by redefining the couplings of nuclear matrix
elements in the reactor $\nub_e$ production and detection processes.
From Eq.~\eqref{eq:normalize_factor}, $\Nt_{e}$ is related to the
average parts of NSIs, and contributes to the normalization factor of
the reactor antineutrino flux. In this work we take
$\Nt_{e}=1.0$ for simplicity, as we mainly stress on the
difference parts of NSIs and the experimental spectral measurements.

\begin{figure}[h!]
\begin{center}
\vspace{-.9cm}
\includegraphics[width=0.75\textwidth]{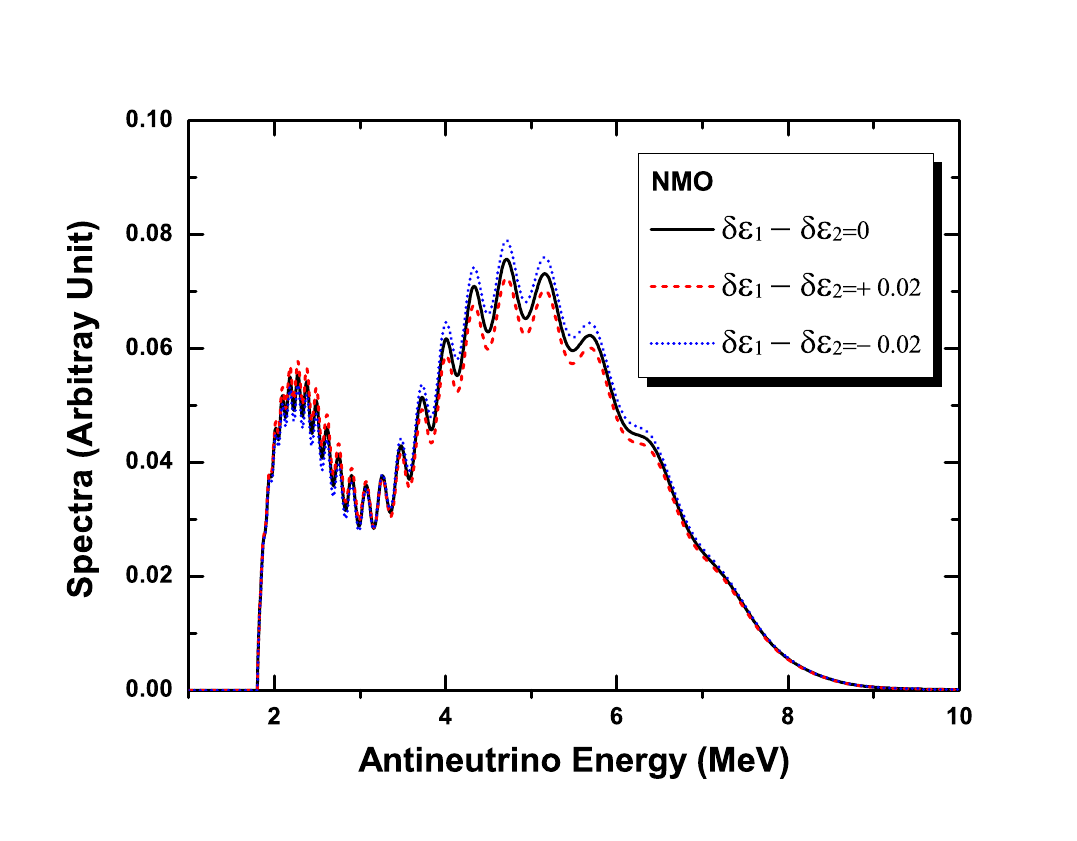}
\vspace{-1.1cm}
\\\includegraphics[width=0.75\textwidth]{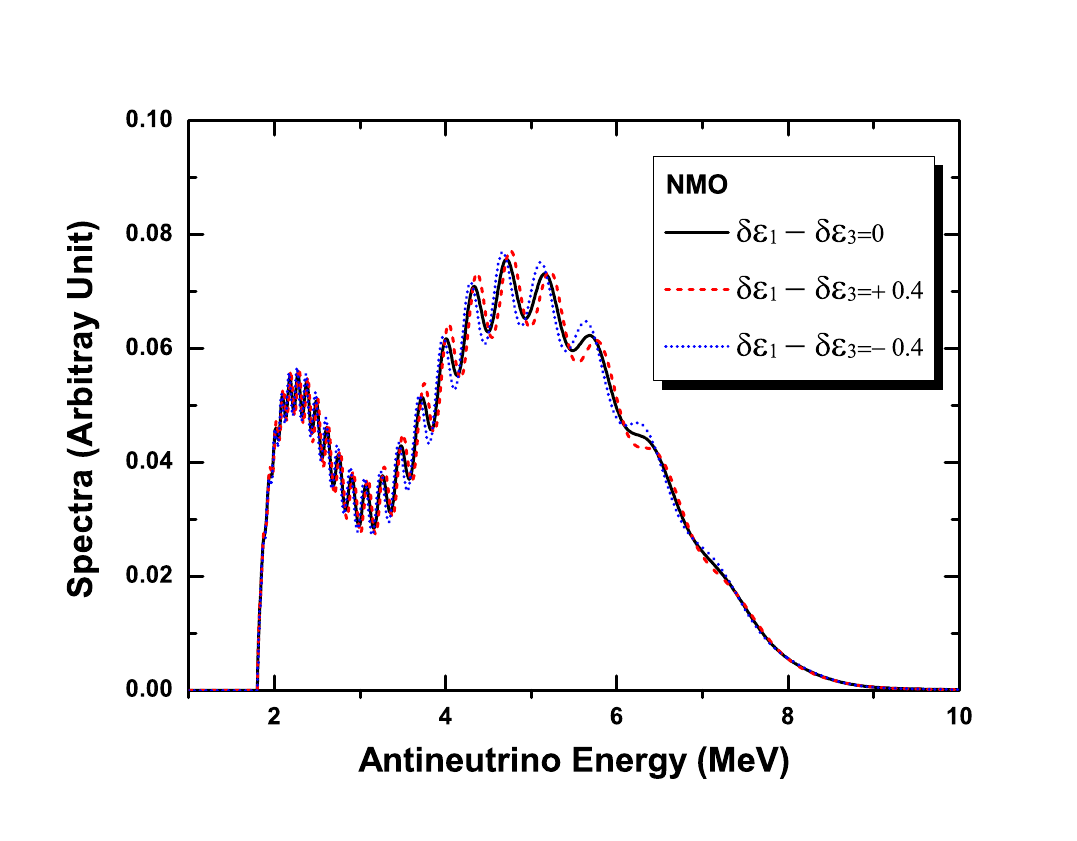}
\vspace{-.8cm} \caption{The effect of NSIs in reactor $\nub_e$
spectra at a baseline of 52.5 km. For visualization,  we set
$\dvarepsilon_1-\dvarepsilon_2=0,\pm0.02$ in the upper panel and
$\dvarepsilon_1-\dvarepsilon_3=0,\pm0.4$ in the lower panel.
$\dvarepsilon_1-\dvarepsilon_3$ is fixed at zero in the upper panel,
and $\dvarepsilon_1-\dvarepsilon_2$ is fixed at zero in the lower
panel.  The oscillation parameters are taken as in
Eqs.~\eqref{eq:datamixing} and~\eqref{eq:datamass}. The NMO is assumed for illustration.}
\label{fig:spectrum}
\end{center}
\end{figure}

We show the effect of NSIs in the reactor $\nub_e$ spectra at a
baseline of 52.5 km in Fig.~\ref{fig:spectrum}, where the influences
of $\dvarepsilon_1-\dvarepsilon_2$ and
$\dvarepsilon_1-\dvarepsilon_3$ are  presented in the upper panel
and lower panel, respectively. The oscillation parameters are taken
as in Eqs.~(\ref{eq:datamixing}) and~(\ref{eq:datamass}). The
scenario of 3-generation  neutrino oscillations with $\dU=0$ is also
shown for comparison. In the upper panel, we fix
$\dvarepsilon_1-\dvarepsilon_3=0$ and find that non-zero
$\dvarepsilon_1-\dvarepsilon_2$ introduces the spectral distortion
to the slow oscillation term $\Delta_{21}$. For
$\dvarepsilon_1-\dvarepsilon_2=0.02$, the spectrum is suppressed in
the high energy region with $E>3$ MeV and enhanced for the low
energy range 2 MeV $<E<3$ MeV. In comparison, negative
$\dvarepsilon_1-\dvarepsilon_2$ gives the opposite effect on the
spectrum distortion. In the lower panel, we set
$\dvarepsilon_1-\dvarepsilon_2=0$ and observe that
$\dvarepsilon_1-\dvarepsilon_3$ can affect the spectral distribution
for the fast oscillation term $\Delta_{31}$. Non-trivial NSI effect
will contribute a small phase advancement or retardance to the fast
oscillation depending upon the sign of
$\dvarepsilon_1-\dvarepsilon_3$.

\subsection{Statistical analysis}

In this part, we shall implement the standard $\chi^2$ statistical
method to do the numerical analysis with the above setup. A general
$\chi^2$ function using the spectrum calculated in
Eq.~\eqref{eq:spectrum_NSIs} can be defined as
\bq \chi^2=
\sum^{N_\text{bin}}_{i=1}\frac{\big[\tilde{M}_i(p^M,\dvarepsilon^M)-\tilde{T}_i(p^T,\dvarepsilon^T)(1+\sum_k
\alpha_{ik}\epsilon_k)\big]^2}{\tilde{M}_i(p^M,\dvarepsilon^M)}
+\sum_k \frac{\epsilon_k^2}{\sigma_k^2}\,, \label{eq:chi2func} \nq
where $\tilde{M}_i$ and $\tilde{T}_i$ are the measured and predicted
(with oscillation) reactor $\nub_e$ fluxes in the $i$-th energy bin
respectively. The definition of bin sizes is identical to that
assumed in Ref.~\cite{Li:2013zyd}. The systematic uncertainties $\sigma_k$ together with the
corresponding pull parameters $\epsilon_k$ for
the nominal setups are also the same as those in Ref.~\cite{Li:2013zyd}, which include the
correlated (absolute) reactor uncertainty ($2\%$), the uncorrelated
(relative) reactor uncertainty ($0.8\%$), the flux spectrum
uncertainty ($1\%$) and the detector-related uncertainty ($1\%$).
The sets of $p$ and
$\dvarepsilon$ are for the standard oscillation parameters and NSI
parameters respectively with $p=\{\thetat_{12},\thetat_{13},\Delta
m^2_{12}, \Delta m^2_{ee}\}$ and, $\dvarepsilon =
\{\dU/{\cost_{12}\cost_{13}}, \pm\dU/{\sint_{12}\cost_{13}},
\pm\dU/{\sint_{13}}\}$ for specified models, or $\dvarepsilon
=\{\dvarepsilon_1-\dvarepsilon_2, \dvarepsilon_1-\dvarepsilon_3 \}$
for the general model defined in section~\ref{subsec:setups}.

\subsubsection{Shifts of mass-squared differences}
Neglecting the existing non-zero NSIs, we may get biased best-fit
oscillation parameters. In this part we shall evaluate the sizes and
properties on the shifts of mass-squared differences due to the NSI
effect. {In the numerical simulation, we use the spectrum with
non-zero NSIs as the true spectrum, and the spectrum of the standard
neutrino oscillation without NSIs as the predicted spectrum. In other
word, the true spectrum is defined in Eq.~\eqref{eq:spectrum_NSIs}
with the oscillation probability given in Eq.
\eqref{eq:reactorprobability}, and the predicted spectrum is given
in Eq.~\eqref{eq:spectrum} with the oscillation probability in Eq.
~\eqref{eq:PeeStandard}. }Then we minimize the $\chi^2$ function and
find out the best-fit mixing angles and mass-squared differences.

\begin{figure}[t!]
\begin{minipage}[t]{0.5\linewidth}
\centering
\includegraphics[width=1\textwidth]{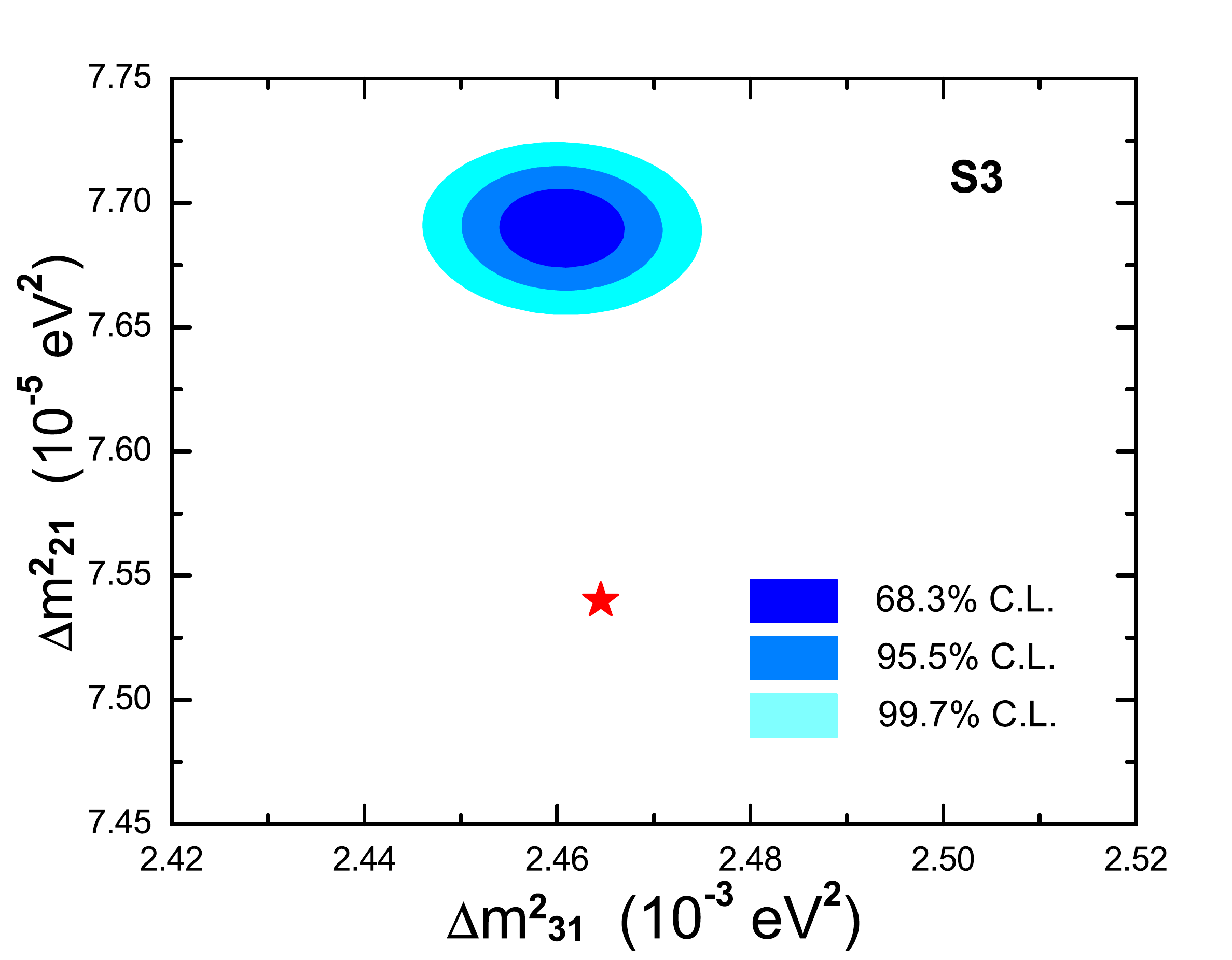}
\includegraphics[width=1\textwidth]{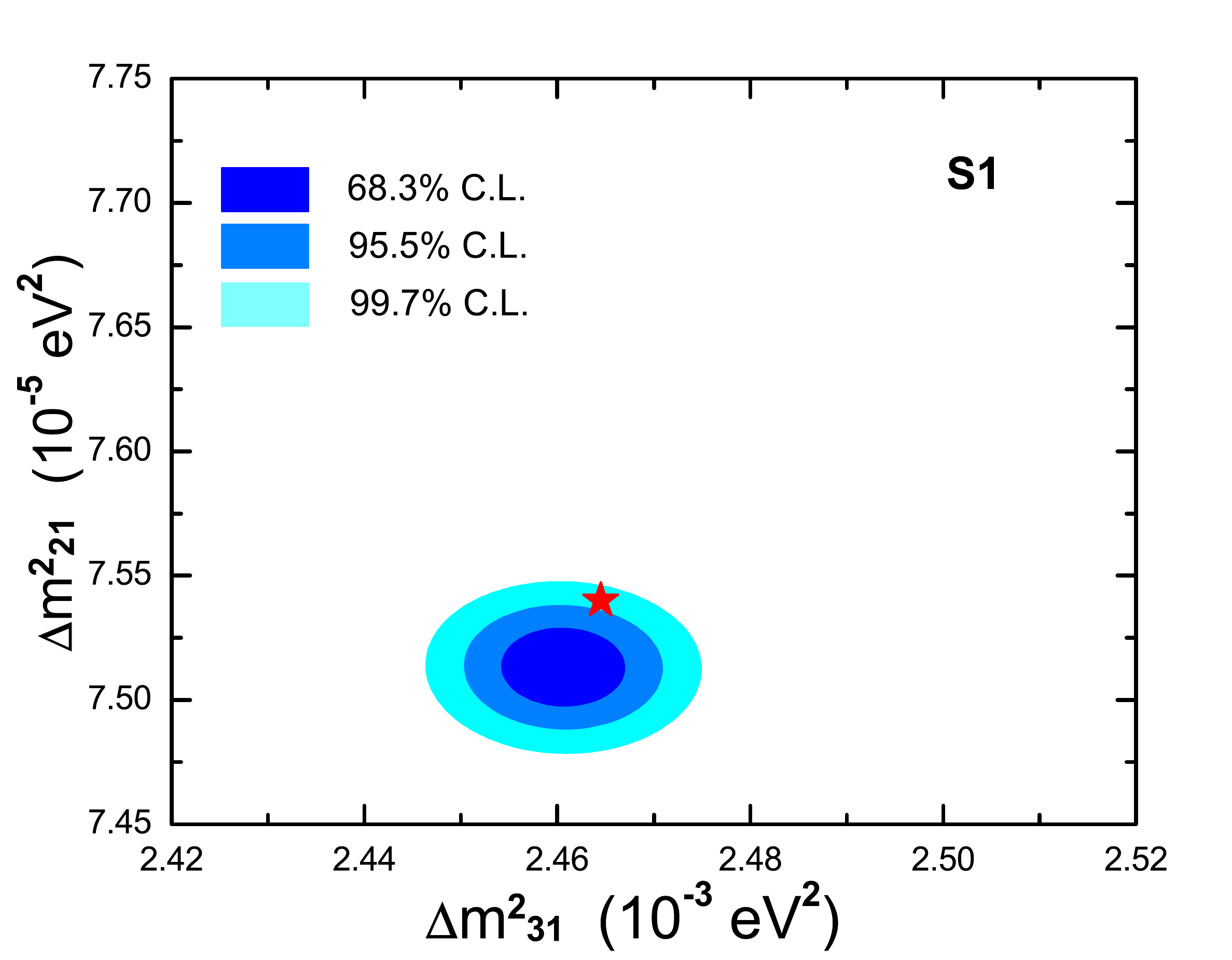}
\end{minipage}
\begin{minipage}[t]{0.5\linewidth}
\centering
\includegraphics[width=1\textwidth]{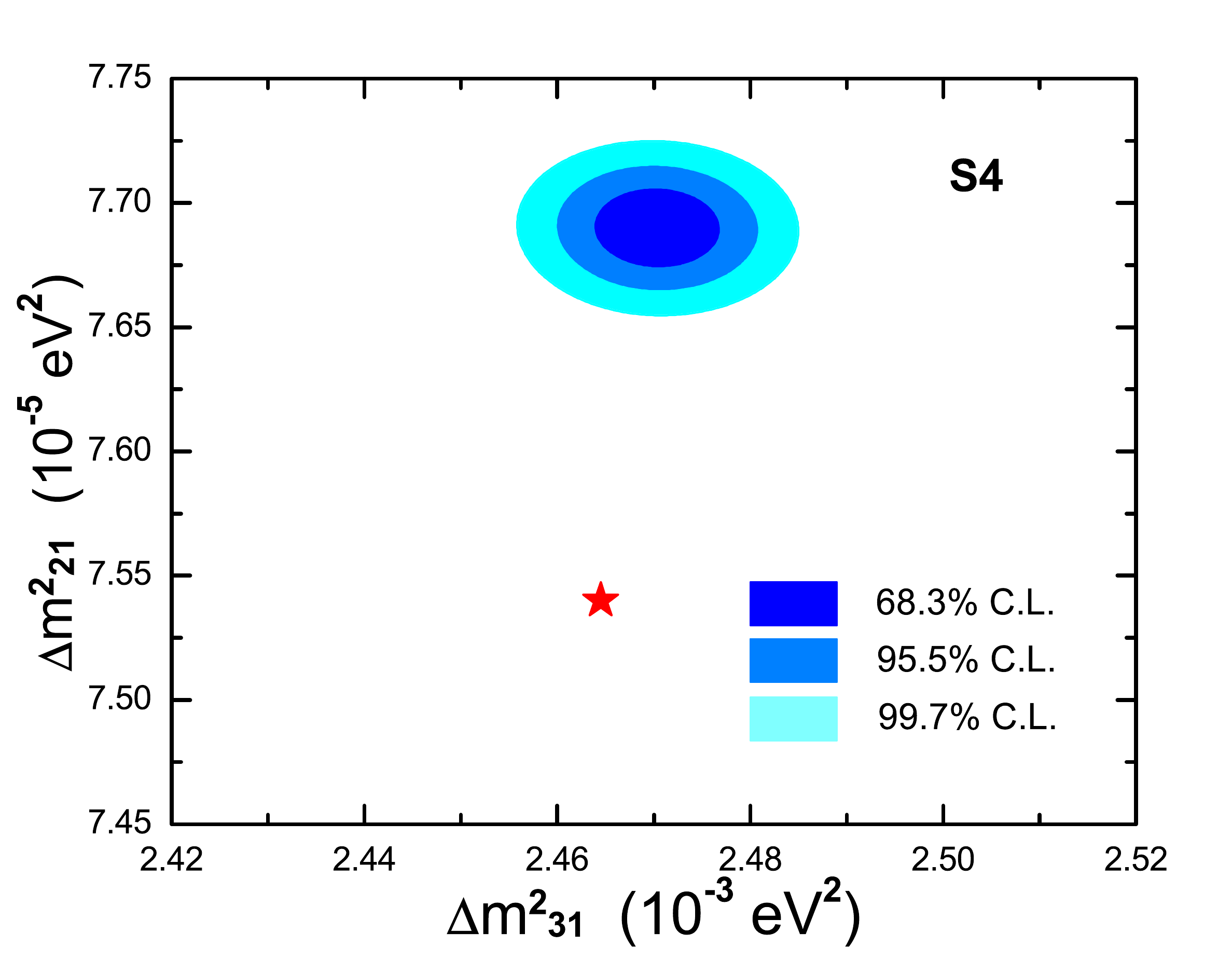}
\includegraphics[width=1\textwidth]{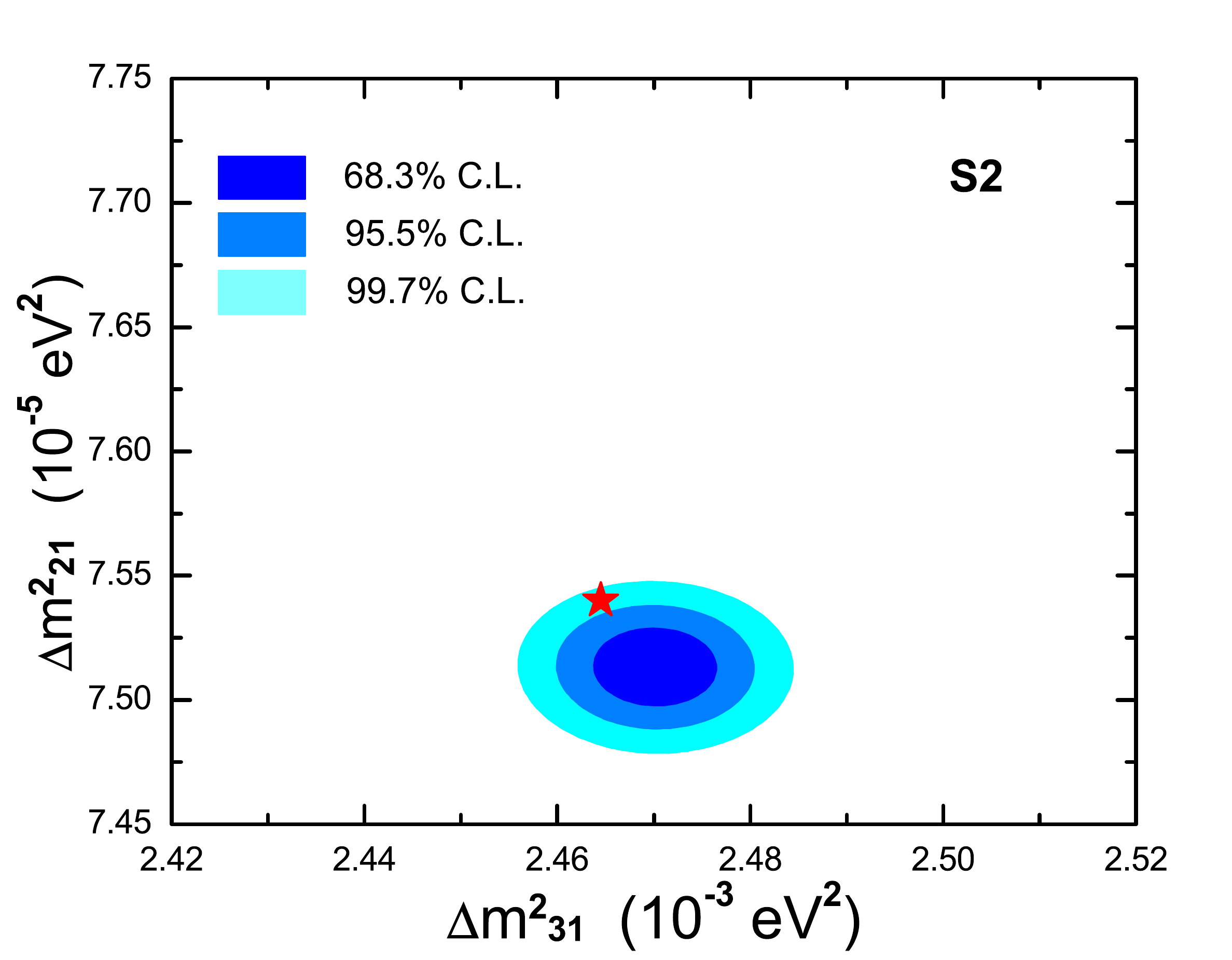}\end{minipage}
\caption{The NSI-induced shifts for $\Delta
{m}^2_{21}$ and $\Delta{m}^2_{31}$ in the four specific models
defined in Eq.~(\ref{eq:specmod}) with $\dU = 0.01$. The red stars
stand for the true values of $\Delta{m}^2_{21}$ and
$\Delta{m}^2_{31}$, and contours are the 68.3\% (1$\sigma$), 95.5\%
(2$\sigma$), 99.7\% (3$\sigma$) allowed regions for $\Delta
{m}^2_{21}$ and $\Delta{m}^2_{31}$ when the NSI effect is neglected.
The NMO is assumed for illustration.}
\label{fig:mass_shift1}
\end{figure}
\begin{figure}[ht!]
\includegraphics[width=0.5\textwidth]{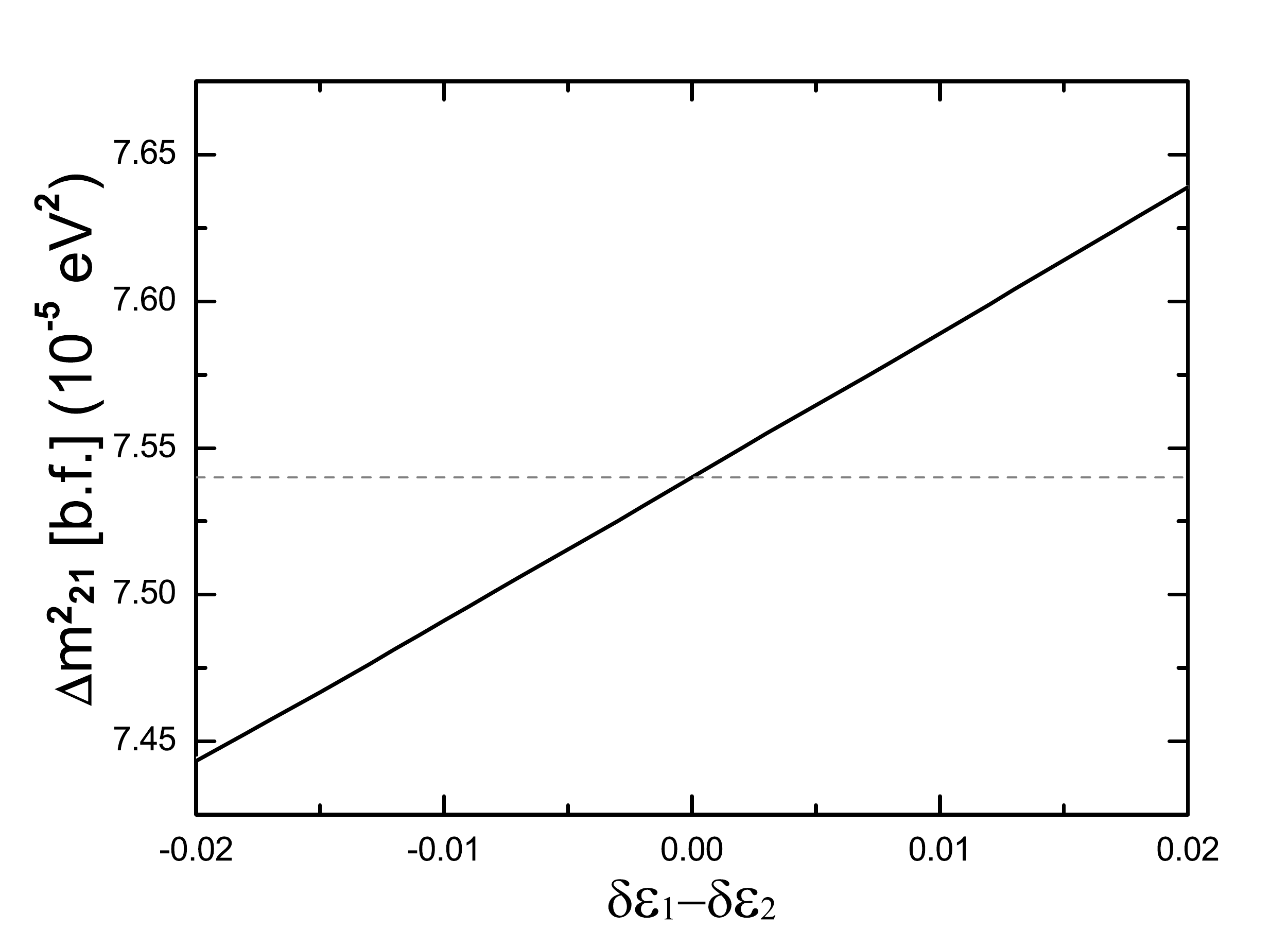}
\includegraphics[width=0.5\textwidth]{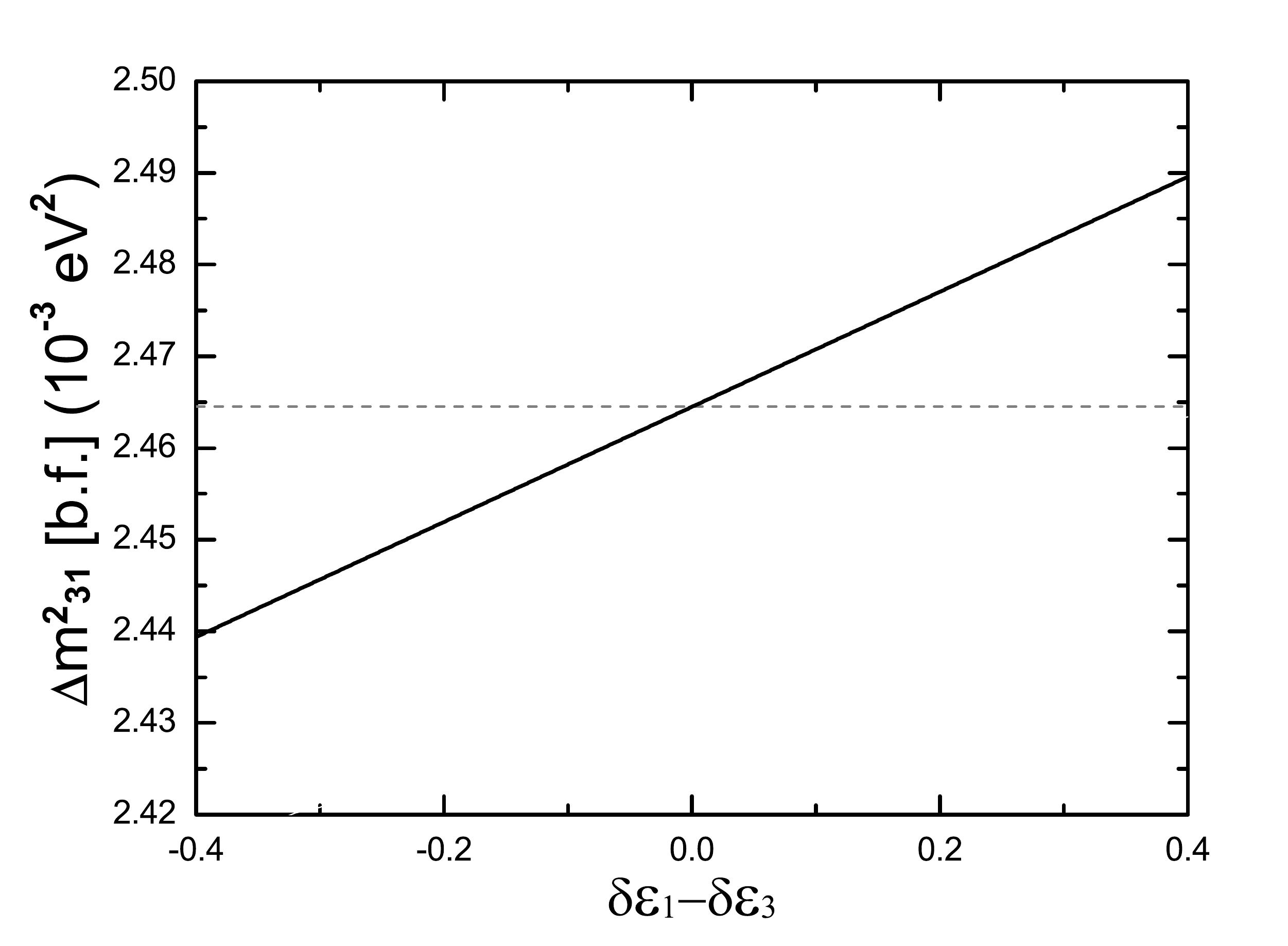}
\caption{The best-fit (b.f.) mass-squared differences
for $\Delta m^2_{21}$ (left panel) and $\Delta m^2_{31}$ (right
panel) as the functions of the true values of
$\dvarepsilon_1-\dvarepsilon_2$ or $\dvarepsilon_1-\dvarepsilon_3$
respectively in the generic treatment of the NSI parameters. The
best-fit values of $\Delta m^2_{21}$ and $\Delta m^2_{31}$ are
obtained by the minimization of the $\chi^2$ function without the
NSI effect. The horizontal dashed lines are for true values of the
mass-squared differences, and the NMO is assumed for
illustration.}\label{fig:mass_shift2}
\end{figure}

The effects of mass shifts are shown in Figs.~\ref{fig:mass_shift1} and \ref{fig:mass_shift2}, corresponding to
the first and second treatments, respectively, where NMO has been
assumed. In the first treatment, the mass-squared differences have
different shift sizes and directions in each specific models,
dependent upon the sign of $\delta U$. The best-fit value of $\Delta
m^2_{21}$ decreases from its true value in the models of S1, S2 or
increases for S3, S4, and the best-fit value of $\Delta m^2_{31}$
decreases in S1, S3 or increases in S2, S4. A simple explanation can
be found in the following estimation. The NSI parameters in these
models are numerically given by
\bq
&&{\rm S1}: \dvarepsilon_1-\dvarepsilon_2=-0.54\,\dU,\quad \dvarepsilon_1-\dvarepsilon_3=-5.60\,\dU, \nonumber\\
&&{\rm S2}: \dvarepsilon_1-\dvarepsilon_2=-0.54\,\dU,\quad \dvarepsilon_1-\dvarepsilon_3=+8.06\,\dU, \nonumber\\
&&{\rm S3}: \dvarepsilon_1-\dvarepsilon_2=+3.00\,\dU,\quad \dvarepsilon_1-\dvarepsilon_3=-5.60\,\dU, \nonumber\\
&&{\rm S4}: \dvarepsilon_1-\dvarepsilon_2=+3.00\,\dU,\quad
\dvarepsilon_1-\dvarepsilon_3=+8.06\,\dU, \nq where $\dU$ is fixed
at 0.01 in Fig.~\ref{fig:mass_shift1}. The sign of
$\dvarepsilon_1-\dvarepsilon_2$ is ``$-$'' in S1, S2 and ``$+$'' in
S3, S4, which reduces or increases the measured value of $\Delta
\tilde{m}^2_{21}$ in the LHS of Eq.~\eqref{eq:mass_shift},
respectively. Similar analysis is also valid for explaining the
shift of $\Delta m^2_{31}$ as shown in Fig.~\ref{fig:mass_shift1}.
Moreover, due to the smallness of the coefficients of
$\dvarepsilon_1-\dvarepsilon_2$ in models S1, S2, the shift of
$\Delta m^2_{21}$ in S1, S2 is much smaller than that in S3, S4. The
relative mass shift for $\Delta m^2_{21}$ is around 0.4\% in S1, S2
and 2\% in S3, S4. Although the magnitude of
$\dvarepsilon_1-\dvarepsilon_3$ is in general larger than
$\dvarepsilon_1-\dvarepsilon_2$, the absolute value of $\Delta
m^2_{31}$ is much larger than $\Delta m^2_{21}$, and thus the
relative shift of $\Delta m^2_{31}$ is not significant, just roughly
around 0.2\% for the best-fit data in four models.

The effects of NSI-induced mass shifts in the general case are
presented in Fig.~\ref{fig:mass_shift2}, without any assumptions on
the relation of NSI parameters. Our simulation results can be
understood using the relation in Eq.~\eqref{eq:mass_shift} that the
fitted mass-squared differences $\Delta m^{2\,\text{J}}_{21}$ and
$\Delta m^{2\,\text{J}}_{31}$ in JUNO are linearly dependent upon
$\dvarepsilon_1-\dvarepsilon_2$ and $\dvarepsilon_1-\dvarepsilon_3$,
respectively. With the JUNO nominal setup, we can simplify Eq.
\eqref{eq:mass_shift} into the following formulae \bq \Delta
m^{2\,\text{J}}_{21}=\Delta
\tilde{m}^{2}_{21}((E/L)^{\text{J}})=\Delta
m^2_{21}+(\dvarepsilon_1-\dvarepsilon_2)4(E/L)^{\text{J}} \nonumber\\
\Delta m^{2\,\text{J}}_{31}=\Delta \tilde{m}^{2}_{31}((E/L)^{\text{J}})=\Delta
m^2_{31}+(\dvarepsilon_1-\dvarepsilon_3)4(E/L)^{\text{J}}
\nq
where
$4(E/L)^{\text{J}}\simeq5\times10^{-5}~\text{eV}^2$ roughly holds.
The relative mass shift of $\Delta m^2_{21}$ is about $\frac{2}{3}
(\dvarepsilon_1-\dvarepsilon_2)$, in the same level of
$\dvarepsilon_1-\dvarepsilon_2$, i.e., in the same order as
$\text{Im}(\dU_{ei})$ and $\depsilon_{\alpha\beta}$. The relative
mass shift of $\Delta m^2_{31}$ is about $0.02
(\dvarepsilon_1-\dvarepsilon_3)$. Keeping in mind 
$\dvarepsilon_3=\text{Im}(\dU_{ei})/\sint_{13}$, we obtain that the relative mass
shift of $\Delta m^2_{31}$ is one order smaller than
$\text{Im}(\dU_{ei})$ and $\depsilon_{\alpha\beta}$. In the case of
the IMO, as $\Delta m^2_{31}=-|\Delta m^2_{31}|$ holds, the shift of
$|\Delta m^2_{31}|$ will go in the opposite direction to the
NMO.

\subsubsection{Impacts on the MO measurement}

\begin{figure}[h!]
\begin{center}\hspace{-0.3cm}
\includegraphics[width=0.71\textwidth]{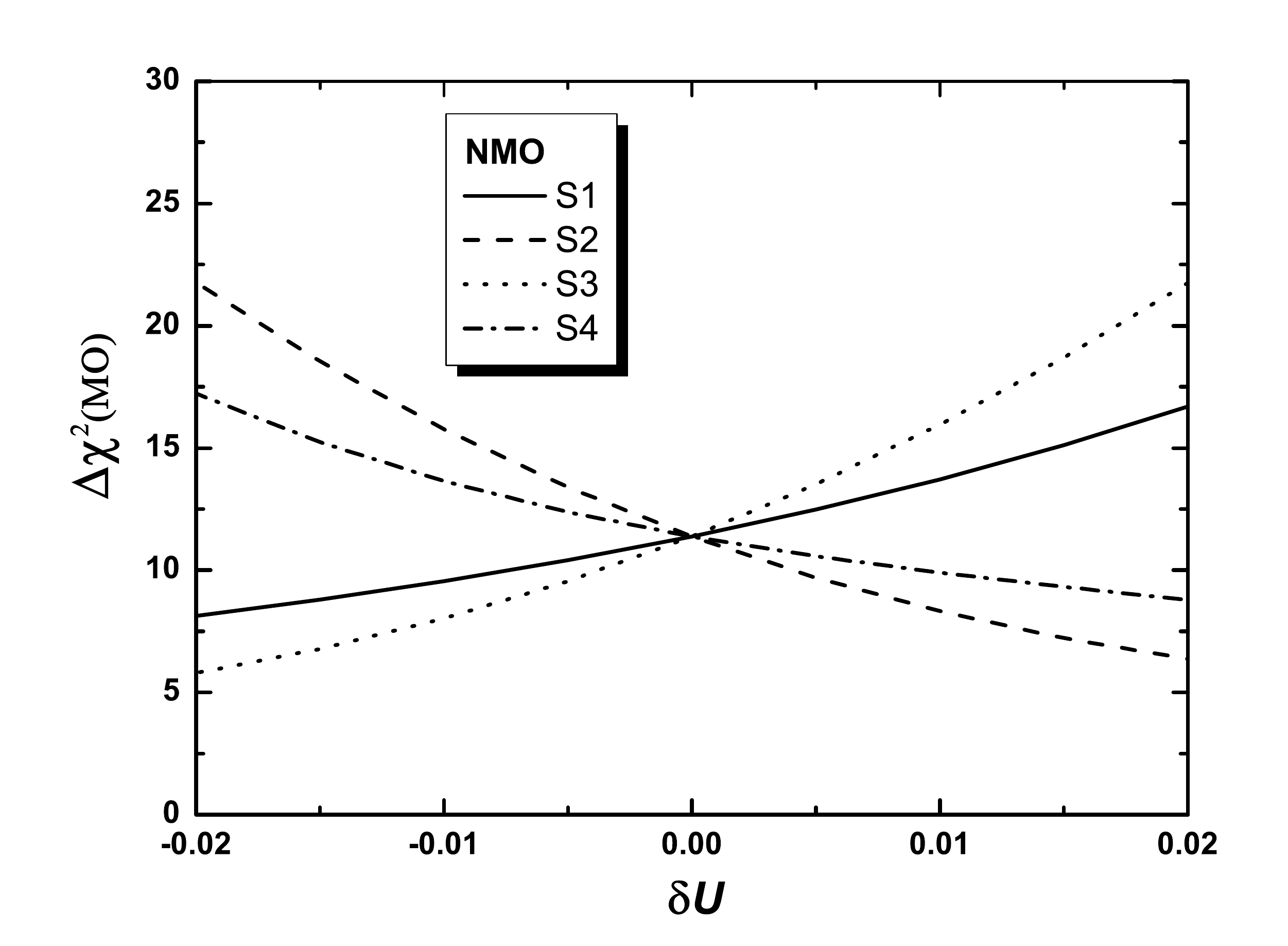}
\vspace{-0.5cm}
\caption{The MO sensitivity for different true values
of the NSI parameter $\delta U$ in the four different specific
models defined in Eq.~(\ref{eq:specmod}). The NMO is assumed for
illustration.}
\label{fig:mh_special}
\end{center}
\end{figure}
\begin{figure}[h!]
\begin{center}
\includegraphics[width=0.7\textwidth]{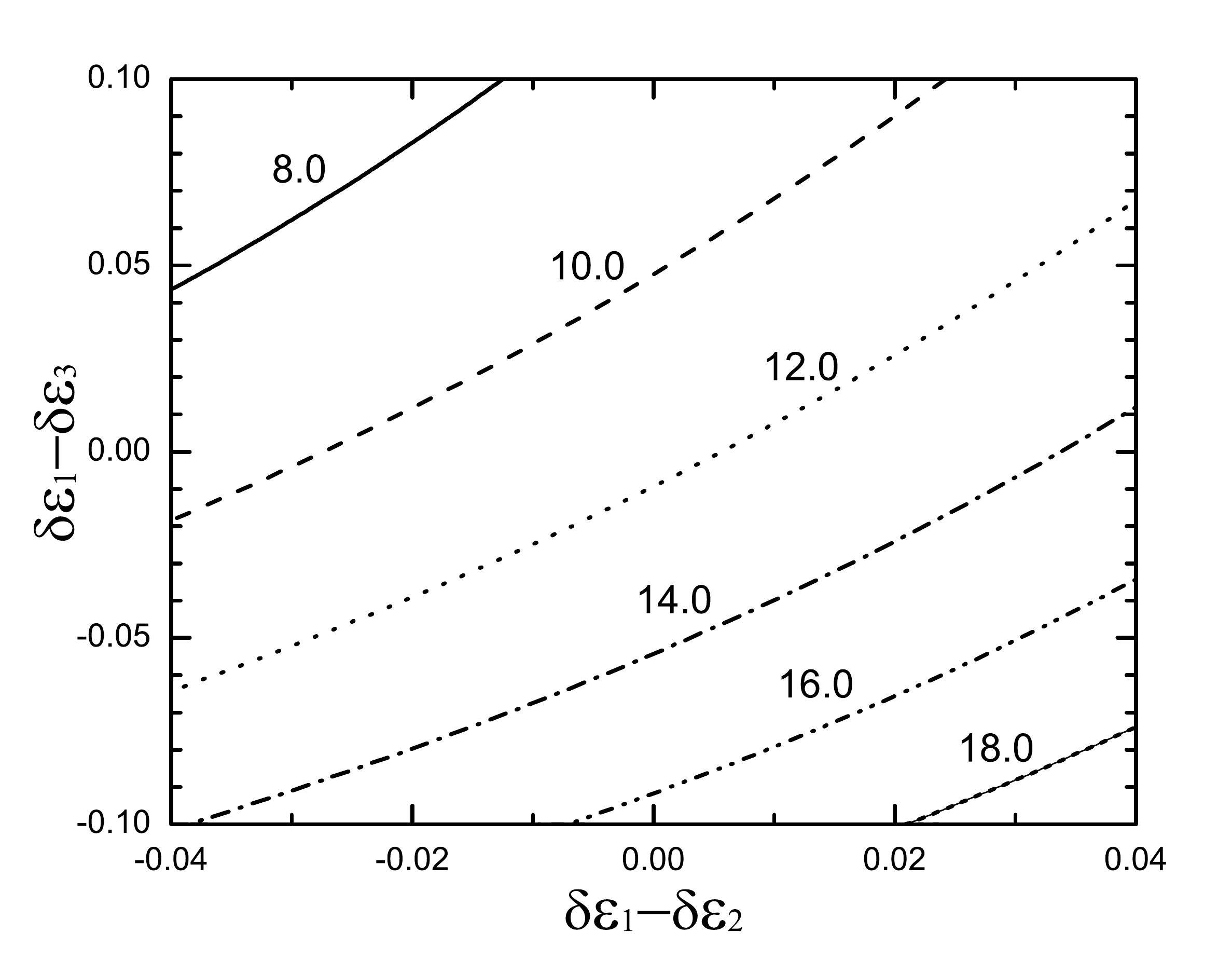}
\vspace{-.5cm} \caption{The iso-$\Delta \chi^2$ contours for the MO
sensitivity in the generic NSI model as a function of two
effective NSI parameters $\dvarepsilon_1-\dvarepsilon_2$ and
$\dvarepsilon_1-\dvarepsilon_3$. The NMO is assumed for
illustration.} \label{fig:mh_iso}
\end{center}
\end{figure}

When fitting the $\chi^2$ function in Eq.~(\ref{eq:chi2func}) with
both NMO and IMO, we can take the difference of the minima to
measure the sensitivity of neutrino mass ordering, where the
discriminator is defined as \bq \Delta
\chi^2(\text{MO})=\big|\chi^2_{\text{min}}(\text{NMO})-\chi^2_{\text{min}}(\text{IMO})\big|\,.
\label{eq:deltachi2}\nq For these specific models defined in
Eq.~(\ref{eq:specmod}), we illustrate in Fig.~\ref{fig:mh_special}
the NSI effect on the MO measurement by showing the dependence of
the MO sensitivity on the true value of the NSI parameter $\delta
U$, where the NMO is assumed for illustration. The NSI effect with a
negative $\dU$ in S1, S3 or positive $\dU$ in S2, S4 will decrease
the $\Delta\chi^2(\text{MO})$ value and thus degrade the sensitivity
of the MO determination. However in the other half possibilities,
the NSI effect can increase the $\Delta\chi^2(\text{MO})$ value and
enhance the MO sensitivity. Moreover, the NSI effect shows stronger
influence on the MO measurement in models S2, S3 than S1, S4.
On the other hand, we illustrate in  Fig.~\ref{fig:mh_iso} the
iso-$\Delta \chi^2$ contours for the MO sensitivity in the generic
NSI model as a function of two effective NSI parameters
$\dvarepsilon_1-\dvarepsilon_2$ and $\dvarepsilon_1-\dvarepsilon_3$.
The NMO is assumed for illustration. We can learn from the figure
that the smaller $\dvarepsilon_1-\dvarepsilon_2$ and larger
$\dvarepsilon_1-\dvarepsilon_3$ will reduce the possibility of the
MO measurement. If $\dvarepsilon_1-\dvarepsilon_2$ decreases by 0.03
or $\dvarepsilon_1-\dvarepsilon_3$ increases by 0.05, $\Delta
\chi^2$ will be suppressed by 2 units.

\subsubsection{Constraints on NSI parameters}

\begin{figure}[h]
\begin{center}
\includegraphics[width=0.75\textwidth]{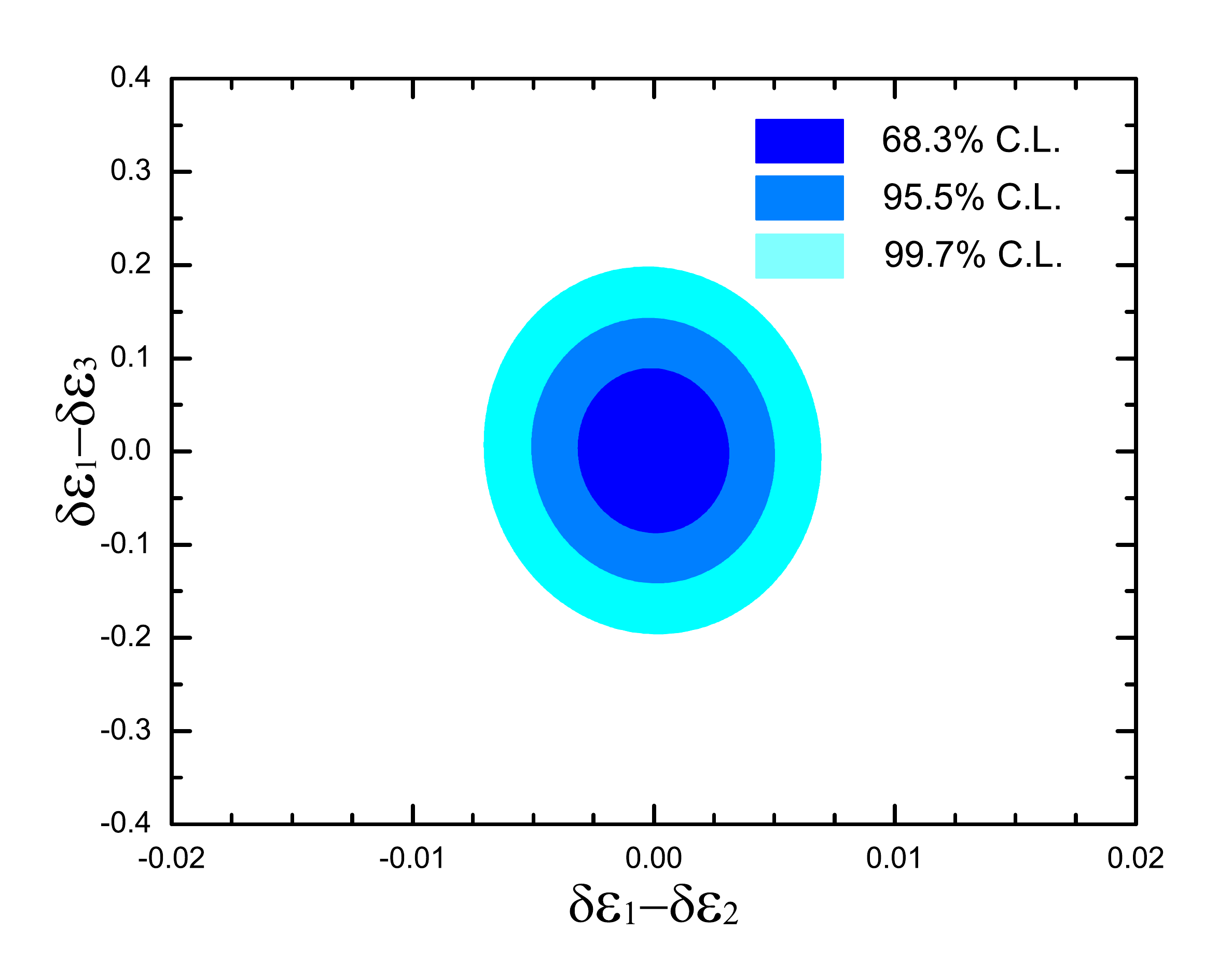}
\vspace{-.5cm} \caption{The experimental constraints on the generic
NSI parameters $\dvarepsilon_1-\dvarepsilon_2$ and
$\dvarepsilon_1-\dvarepsilon_3$, where the true values are fixed at
$\dvarepsilon_1-\dvarepsilon_2=\dvarepsilon_1-\dvarepsilon_3=0$, and
the contours are the 68.3\% (1$\sigma$), 95.5\% (2$\sigma$), 99.7\%
(3$\sigma$) allowed regions. The NMO is assumed for illustration.}
\label{fig:paralimit}
\end{center}
\end{figure}

In this part we shall discuss the constraints on the NSI parameters
with the JUNO nominal setup. In our numerical calculation,
the true oscillation parameters are taken as in
Eqs.~(\ref{eq:datamixing}) and~(\ref{eq:datamass}), and the true NSI
parameters are taken as
$\dvarepsilon_1-\dvarepsilon_2=\dvarepsilon_1-\dvarepsilon_3=0$. In
the fitting process, we fix the oscillation parameters but take
the NSI parameters as free. With the above simplification, we can
obtain the constraints on the considered NSI parameters. In
Fig.~\ref{fig:paralimit}, we show the limit on these two parameters
at the 1, 2, 3$\sigma$ confidence levels. For
$\dvarepsilon_1-\dvarepsilon_2$, the precision is much better than 1\%.
However, the precision for $\dvarepsilon_1-\dvarepsilon_3$ is around
the $10\%$ level. JUNO is designed for a precision spectral measurement at the
oscillation maximum of $\Delta m^2_{21}$. From
Eq.~(\ref{eq:reactorprobability}), the precision for
$\dvarepsilon_1-\dvarepsilon_2$ can be compatible with that of
$\sin^2 2\theta_{12}$, 
where a sub-percent level can be achieved \cite{Li:2014qca}. On the other hand, the
precision for $\sin^2 2\theta_{13}$ is also at the $10\%$ level,
also consistent with that of $\dvarepsilon_1-\dvarepsilon_3$ in our
numerical simulation. Because $\dvarepsilon_1-\dvarepsilon_3$ is
suppressed by $\sin\thetat_{13}$, the above two constraints are
actually compatible if we consider the physical NSI parameters
$\depsilon_{\alpha\beta}$ defined in Eq.~(\ref{eq:definition}).
Notice that different assumptions (e.g., the uncertainties of
oscillation parameters) on the experimental systematics may alter
the quantitative precision of the NSI parameters, but our
qualitative conclusion is reasonable in any realistic systematical
assumptions.

\section{Conclusion}

In this work we have presented a complete and new derivation on the
generic NSI effects in reactor antineutrino oscillations, where the
NSI parameters are divided into the average and difference parts of
the antineutrino production and detection processes. The average
part can induce an effective non-unitary PMNS matrix and shift the
true values of the mixing angles. On the other hand, the difference
part of the NSI effect can be parametrized with only two independent
parameters (i.e., $\dvarepsilon_1-\dvarepsilon_2$,
$\dvarepsilon_1-\dvarepsilon_3$), and give the energy- and
baseline-dependent corrections to the mass-squared differences. Eq.
\eqref{eq:reactorprobability} is our key formula for the reactor
$\nub_e\to\nub_e$ survival probability, where
\begin{itemize}
\item we define the mixing angle shifts as the deviations of measured mixing angles
$\thetat_{12}$, $\thetat_{13}$ from their true values $\theta_{12}$,
$\theta_{13}$. However, we stress that these constant shifts are
undetectable in reactor antineutrino experiments.

\item the two NSI parameters $\dvarepsilon_1-\dvarepsilon_2$ and
$\dvarepsilon_1-\dvarepsilon_3$ can be absorbed into the
mass-squared differences and the corresponding $E/L$-dependent
effective parameters $\Delta \tilde{m}^2_{12}$ and $\Delta
\tilde{m}^2_{31}$ can be defined as the shifts of mass-squared
differences. These shifts are detectable in the spectral
measurement of reactor antineutrino oscillations. 
\end{itemize}

Our analytical formalism is applied to the future medium baseline
reactor antineutrino experiment JUNO. Two different treatments (a
class of specific models and the most general case with the full
parameter space) of the NSI parameters are employed in our numerical
analysis. We analyze the NSI impact on the precision measurement of
mass-squared differences and the determination of the neutrino mass
ordering, and present the JUNO sensitivity of the relevant NSI
parameters. Numerically,
\begin{itemize}
\item we find that the relative mass shift of $\Delta m^2_{21}$
is around $\frac{2}{3}(\dvarepsilon_1-\dvarepsilon_2)$, in the same
order of the original NSI parameters $\depsilon_{\alpha\beta}$; and
the relative shift of $\Delta m^2_{31}$ is around
$0.02(\dvarepsilon_1-\dvarepsilon_3)$, one order smaller than the
magnitude of $\depsilon_{\alpha\beta}$. However, cancelations may
appear in $\dvarepsilon_1-\dvarepsilon_2$ and suppress the mass
shift of $\Delta m^2_{21}$ (see the models S1 and S2).

\item a positive $\dvarepsilon_1-\dvarepsilon_2$ or negative
$\dvarepsilon_1-\dvarepsilon_3$ may enhance the sensitivity of the
neutrino MO measurement at JUNO.

\item due to the specific configuration of JUNO, the constraint on $\dvarepsilon_1-\dvarepsilon_2$ can be
better than 1\%, but $\dvarepsilon_1-\dvarepsilon_3$ can only be
measured at the 10\% precision level.

\end{itemize}

Compared with long baseline and short baseline reactor antineutrino
experiments (e.g., KamLAND and Daya Bay), the medium baseline
reactor antineutrino experiment JUNO is more suitable for
constraining the NSI effect because both the slow and fast
oscillation terms are measurable in the reactor antineutrino
spectral measurement. Taking into account the complementary
properties of reactor antineutrino experiments at different
baselines, it is desirable to present a sophisticated global
analysis of all the reactor antineutrino experiments and therefore,
we may obtain the most complete and precision testing of the NSIs or
other new physics beyond the SM.

\section*{Acknowledgements}

We are indebted to Z.Z. Xing for his continuous encouragement and
reading the manuscript. We are also grateful to S. Zhou for useful
discussions. This work was supported in part by the National Natural
Science Foundation of China under Grant Nos. 11135009, 11305193, and
in part by the Strategic Priority Research Program of the Chinese
Academy of Sciences under Grant No. XDA10010100.

\end{document}